\documentclass[journal]{IEEEtran}
\usepackage{array}
\usepackage{graphicx}
\graphicspath{{figures/}}
\usepackage{float} 
\usepackage{cite} 
\usepackage{amsmath,amssymb,amsfonts,bm}
\usepackage[hyphens,spaces,obeyspaces]{url} 
\usepackage{circuitikz}
\usepackage{mathrsfs}   
\usepackage{amsmath}
\usepackage{pgfplots} 
\pgfplotsset{compat=1.3, tick style={black}}
\usepackage{multirow}
\usepackage{xcolor}
\usepackage[font=footnotesize,caption=false,subrefformat=parens,labelformat=parens]{subfig}
\usepackage{algorithm}
\usepackage{algorithmic}
\usepackage{etoolbox}
\usepackage[colorinlistoftodos, textsize=small]{todonotes}
\usepackage{soul}
\usepackage{hyperref}
\hypersetup{
    colorlinks=true,
    linkcolor=blue,     
    urlcolor=cyan,
    }
\definecolor{LightCyan}{rgb}{0.88,1,1}

\ifCLASSINFOpdf
\else
\fi

\title{Data-driven, Internet-inspired, and Scalable EV Charging for Power Distribution Grid}

\author{\IEEEauthorblockN{Emin Ucer\IEEEauthorrefmark{1},  Mithat Kisacikoglu\IEEEauthorrefmark{1}, Murat Yuksel\IEEEauthorrefmark{2}, and Ali C. Gurbuz\IEEEauthorrefmark{3}}
\\
\IEEEauthorblockA{\IEEEauthorrefmark{1} Dept. of Electrical and Computer Engineering,
University of Alabama, Tuscaloosa, AL}\\
\IEEEauthorblockA{\IEEEauthorrefmark{2} Dept. of Electrical and Computer Engineering Department, University of Central Florida, Orlando, FL}\\
{\IEEEauthorrefmark{3} Dept. of Electrical and Compute Engineering, Mississippi State University, Starkville, MS\\ 
E-mails: eucer@crimson.ua.edu, mkisacik@ua.edu, murat.yuksel@ucf.edu, gurbuz@ece.msstate.edu}
 \thanks{This material is based upon work supported by the National Science Foundation under Award No~1755996.}
 }
\begin{document}


\IEEEoverridecommandlockouts

\IEEEpubid{\begin{minipage}[t]{\textwidth}\ \\[10pt]
        \centering\footnotesize
        {~\copyright2022 IEEE Personal use of this material is permitted.  Permission from IEEE must be obtained for all other uses, in any current or future media, including reprinting/republishing this material for advertising or promotional purposes, creating new collective works, for resale or redistribution to servers or lists, or reuse of any copyrighted component of this work in other works.}
\end{minipage}} 


\maketitle

\begin{abstract}
Electric vehicles (EVs) are finally making their way onto the roads. However, the challenges concerning their long charging times and their impact on congestion of the power distribution grid are still waiting to be resolved. With historical measurement data, EV chargers can take better-informed actions while staying mostly off-line. Proposed solutions that depend on heavy communication and rigorous computation for optimal operation are not scalable. The solutions that do not depend on power distribution topology information, such as Droop control, are more practical as they only use local measurements. However, they result in sub-optimal operation due to a lack of a feedback mechanism. This study develops a distributed and data-driven congestion detection methodology embedded in the Additive Increase Multiplicative Decrease (AIMD) algorithm to control mass EV charging in a distribution grid. The proposed distributed AIMD algorithm performs very closely to the ideal AIMD regarding fairness and congestion handling. Its communication need is almost as low as the Droop control. The results can provide crucial insights on how we can use data to reveal the inner dynamics and structure of the power grid and help develop more advanced data-driven algorithms for grid-integrated power electronics control.
\end{abstract}


% \fi
\IEEEpeerreviewmaketitle

\vspace{-2mm}
\section{Introduction}

\IEEEPARstart{H}{igh} penetration of electric vehicles (EVs) with uncontrolled charging will cause transformer and/or line congestion in the power distribution grid. Among some adverse effects of this congestion are severe voltage drops, increased peak loading, thermal overheating, and even failure of equipment~\cite{erden2015examination,fernandez2011assessment,shafiee2013investigating,veldman2015distribution,leemput2014impact,Clement2009}. Therefore, control of EV charging has become an important research effort to mitigate the impacts. Demand-side load management is a technique that is used to modify customer demand through various tools and methods. Curtailing this demand at peak hours is known as \textit{peak-shaving}. It is used to eliminate short-term demand spikes by lowering and smoothing out peak loads; this prevents equipment overloading. In this study, we will investigate a peak-shaving methodology to reduce the substation peak loading caused by massive EV integration. 

Commonly proposed methods for EV charging control require excessive system information, e.g., the grid topology, load forecasting, and customer preferences~\cite{Richardson2012Optimal,Sojoudi2011Optimal,Restrepo2018Three,Zhang2017Scalable,Liu2017Electric}. An uninterrupted connectivity is also required to communicate these information to the EV chargers and send/receive control commands to/from the chargers. Assuming these are available, various optimization problems have been formulated to achieve certain objectives (e.g. minimizing generation cost, losses and peak load, maximizing capacity utilization and total charging power) while respecting system constraints such as voltage limits and equipment overloading. Richardson et al. use linear programming to maximize the total EV charging power while operating within the grid network's limits~\cite{Richardson2012Optimal}. Sojoudi and Low solve an optimal power flow problem to minimize the generation and charging costs while satisfying the network constraints~\cite{Sojoudi2011Optimal}. Restrepo et al. propose a three-stage control algorithm by solving an optimization problem in each stage to coordinate EV load under volt-var control devices~\cite{Restrepo2018Three}. Zhang et al. formulate the network-constrained EV charging control as a convex problem based on a linearized distribution grid model~\cite{Zhang2017Scalable}. Liu et al. propose a valley-filling solution for EV charging as an optimization problem~\cite{Liu2017Electric}.

A traditional approach is to form an optimal power flow (OPF) problem and deterministically solve it using forecasted generation and demand values to determine the control set points. Recently, machine learning (ML) is also used to develop control algorithms that can evolve behaviors based on OPF simulations~\cite{Ibrahim2020Machine}. The idea is to learn local controller rules from a dataset generated by OPF solutions. In general, performing these optimization tasks is very challenging since the required system model and inputs are either missing or very hard to obtain completely. In \cite{Karagiannopoulos2019Datadriven} and  \cite{Dobbe2020Toward}, data-driven local controllers are proposed by solving an off-line OPF problem using ML regression models. Since these controllers are trained based on a historical and pre-defined dataset, they are vulnerable to uncertainties not covered in the dataset, and instances caused by policy and natural shifts~\cite{Dobbe2020Learning}.

A secondary low-voltage (LV) network is a part of power distribution grid which carries electric energy from distribution transformers to service meters of end customers. Grid topology, especially in LV distribution grids, is often unknown and not available~\cite{Weckx2015Voltage}. Besides, the topology information is usually limited to primary network. The lengths (impedance) of the underground cables connected between houses and distribution feeder are often not available. Moreover, the information regarding phase connections are also often unknown~\cite{Weckx2015Voltage}. Previously mentioned solutions heavily depend on this information to perform as desired.

On the other hand, model-free approaches are generic and not dependent on the OPF model, i.e., system topology and variables. They usually use measurements (voltage, frequency, etc.) or historical data to obtain information regarding the grid's dynamic behavior. In this line of work, in a previous study, we presented a decentralized EV charging solution that uses the local voltage measurements and adapts the Internet's TCP/IP protocol for congestion detection~\cite{Ucer2020Decentralized}. Xia et al. proposed decentralized EV charging solutions using local and historical measurements~\cite{Xia2015Local,Xia2014electric}. Geth et al. examine the adaptation of droop curve control for EV load management and their impacts on voltages~\cite{Geth2012Voltage}. Although these solutions are communication free, simpler, and faster, they often try to implicitly detect the peak loading of the grid via measurements and predefined set points and therefore result in sub-optimal control performance.

The Internet uses the Additive Increase and Multiplicative Decrease (AIMD) algorithm for its own congestion management. %Low and
%Lapsley showed that this solution can provide
%stable operation and proportional fairness among users~\cite{1999-low-optimization}.
The only needed information for AIMD is a binary (yes/no) feedback that signifies the congestion event (CE)~\cite{1989-chiu-analysis}. In the context of distribution grid, the CE corresponds to the overloading of substation transformer feeder that powers the distribution network. The CE is triggered when the substation is loaded over its capacity. Note that the CE can also be triggered when there is a need to shave the peak loading in a demand side load management scenario. Therefore, AIMD can be used to flatten the substation's load.

The problem with the AIMD algorithm for EV charging is that its ideal implementation in the distribution grid requires explicit feedback of the CE (substation loading). The studies presented in \cite{Studli2012AIMD,studli2013performance,Zishan2021Adaptive} adapted AIMD for EV charging by using an explicit feedback signal of the CE broadcast to them in real-time. This, however, necessitates real-time connectivity for ideal operation. On the other hand, the Internet's current TCP/IP protocol implements the AIMD algorithm based on local measurements, implicitly detecting the congestion without requiring explicit feedback signal or system knowledge. Our previous works \cite{Ucer2020Decentralized,ucer2018analysis,ucer2019internet,ucer2019analysisPESGM} and other studies \cite{xia2014distributed,Zangs2016} investigated implicit ways of estimating the CE in the distribution grid with only local voltage measurements. However, they mostly stay limited and sub-optimal due to a lack of a feedback mechanism.

This study explores how local historical measurement data can be utilized to learn a machine learning (ML) model that can predict the grid's congestion level using local voltage measurements in real-time. We integrate this model into an AIMD-based EV charging control algorithm to improve distributed EV charging control performance. Our methodology is to learn an ML model that maps local voltage to feeder loading and use it as feedback for the AIMD controller to predict the substation load in real-time using local voltage measurements without requiring connectivity to a central controller. Hence, our approach differs from other studies because we do not require system topology and loading information for operation (model-free). We do not need a real-time feedback signal (reduced communication). We also do not learn rules for optimum controller actions using ML.

The proposed solution does not limit the total number of nodes. Therefore, it is scalable and can easily scale up to as many nodes as possible since the algorithm's operation is independent of each node. The only limit we have is the requirement to occasionally get the substation loading information to train the end-node ML models against system changes/updates. The model parameter variances may also increase as the number of end-nodes increases in a larger network.

This study aims to develop a data-driven, local feedback mechanism for the AIMD controller that can estimate the CE offline. This way, our controller can operate more in a decentralized fashion and take advantage of its benefits. We will also provide a comparative study to test our proposed solution and compare it with two other solutions to evaluate the performance better.

Our main contributions are two folds:

\begin{itemize}
\item We devised an EV charging management solution that can work in a decentralized manner using solely end-node voltage without needing any real-time connectivity to receive charging commands.
\item The proposed solution does not require the grid topology or to forecast the load profiles for solving an optimization problem. It uses historical end-node voltage and grid power measurements (i.e., data-driven) to determine critical voltage thresholds for charging control.
\end{itemize}

The rest of the paper is structured as follows:
In Section-\ref{section:aimd}, the AIMD algorithm is introduced and analyzed. Section-\ref{section-3} investigates the relationship between the grid voltage and the demand power. A counterpart AIMD charging algorithm for EV charging is proposed in Section-\ref{sec:proposed_algorithm}.  Section-\ref{section:other_algorithms} presents performance metrics and two published benchmark studies for comparison with the proposed solution. Section-\ref{section:results} presents the results, and Section-\ref{section:conclusions} concludes the paper.

\section{The AIMD Algorithm} \label{section:aimd}

% \begin{figure}[b]
%   \centering
% 	\includegraphics [trim={0cm 0cm 0cm 0cm}, clip, scale=0.27] {aimd}
% 	\vspace{-3mm}
% 	\caption{Capacity share $w(n)$ over time with AIMD in action.}
%     \label{fig:aimd_figure}
%     \vspace{-0.6cm}
% \end{figure}

Today's Internet owes its standards and protocols to years of debates and research around solving the congestion problem. Its early days suffered congestion control challenges as the number of endpoints drastically increased~\cite{1988-jacobson-congestion}. Observed congestion collapses~\cite{2000-floyd-congestion,1984-nagle-congestion} revealed the need for a solution that assures both the stability of the system and the fair and efficient utilization of the network capacity. To that end,
The AIMD algorithm~\cite{1989-chiu-analysis} was proposed as a congestion avoidance solution at the endpoints. %This solution has been proved to be stable for a wide-scale deployment of end-nodes in a complex system, and maintain proportional fairness among users~\cite{1999-low-optimization,ucer2019analysisPESGM}. %This section will present a mathematical overview of the AIMD algorithm; it will provide a general definition, proof of convergence, and steady-state form as a function of its parameters. 
A detailed mathematical analysis of the AIMD algorithm was presented in a previous study~\cite{ucer2019analysisPESGM}.

AIMD takes actions each time a CE occurs on the network. However, there is no central authority on the Internet that detects CEs and regularly transfers a binary CE signal to end-nodes. Instead, the Internet's existing transmission control protocol (TCP) uses a decentralized and indirect way to determine the degree of the congestion. Each end-node infers the congestion level by measuring how long it takes to get the acknowledgement of a transmitted data packet from its destination, i.e., round-trip time (RTT). 
This statistical learning of RTT provides some partial information of the congestion that might exist in the shared network paths. We presented a detailed work on the explanation of this indirect approach and its adaptation for EV charging problem in \cite{Ucer2020Decentralized}.
Although this indirect approach stays sub-optimal, it has been quite successful. It was proved by Low and Lapsley that this straightforward solution can maintain stable operation and proportional fairness among end-nodes~\cite{1999-low-optimization}. From the standpoint of allocating a limited resource among competing participants in a distributed fashion, AIMD stands out as a strong candidate. It can be modified into various fields with similar problems. Corless et al. provides a detailed mathematical analysis and modeling for AIMD and also discusses possible areas of applications~\cite{AIMDbook}.

The AIMD algorithm has two operational phases. The additive increase (AI) phase takes part when there is available capacity in the system. In this phase, agents are allowed to increase their shares linearly by a rate $\alpha>0$. In the case of CE, the algorithm switches to the multiplicative decrease (MD) phase where agents scale down their shares by a factor $0<\beta<1$. The capacity share of an agent at time $t+1$ can be formulated as follows:
\vspace{-1mm}
\begin{equation}\label{eq:AIMD_eq}
  w_{i}(t+1) =
  \begin{cases}
   w_{i}(t) + \alpha_{i} & \text{if there is no congestion} \\
   w_{i}(t) \times \beta_{i} & \text{if congestion occurs} \\

  \end{cases}
  \vspace{-1mm}
\end{equation}
 where $w_{i}$ denotes the share of the agent $i$. %However, this piece-wise formulation is not convenient for a dynamic analysis, and thus we need a proper mathematical model of the system. %There are many approaches studied to model the algorithm. 
\eqref{eq:AIMD_eq} can be modified for a feeder-level peak-shaving problem for EV charging in the distribution grid as:
\begin{equation}\label{eq:AIMD_EV}
  I_{i}(t+1) =
  \begin{cases}
   I_{i}(t) + \alpha_{i} & \text{if peak-shaving is \underline{not needed}} \\
   I_{i}(t) \times \beta_{i} & \text{if peak-shaving is \underline{needed}} \\
  \end{cases}
  \vspace{-1mm}
\end{equation}
where $I_{i}$ is the charging current of the $i^{th}$ EV. CE here refers to overloading of some distribution grid assets,  primarily the substation feeder. The effectiveness of AIMD to avoid a CE depends on how accurately the congestion or loading level is estimated by users.

Similar to the Internet case, there will be no central authority in our AIMD implementation that will notify end-nodes of CEs. Our proposed method will enable end-nodes to implicitly detect congestion based on their local voltage measurements. The proposed approach differs from our previous work \cite{Ucer2020Decentralized} in that we develop a supervised ML model to map local voltage measurements to feeder loading level by using historical data. Thus, we can improve the congestion detection accuracy and get closer to the optimum operation and more efficient capacity utilization. In the following section, we will investigate how the voltage of the network can be used to estimate the loading level implicitly.
\vspace{-0.1cm}
\section{Analysis of Voltage vs. Power Relationship} \label{section-3}
The AIMD algorithm is based on detecting network congestion, which means the network infrastructure is close to being overloaded or it needs to reduce its peak loading. In a distribution system, the substation transformer could be a potential congestion point since it supplies power to the entire downstream network. Notifying the end-nodes of the substation congestion can be performed by a direct communication link, however, this requires an uninterrupted real-time connectivity. It is much less demanding in terms of communication needs if the local end-nodes can detect this congestion. In this regard, local voltage measurements can be an indicator for this congestion information. The voltage varies depending on the loading level on the grid. Therefore, each node can make an estimation of the substation loading by observing their voltage if they have a guideline that maps their voltage level to the total power drawn from the substation. This guideline can be referred as the voltage and power relationship, and it is investigated in various works~\cite{Ganu2013nPlug, Xia2015demand, Ucer2020Machine}.%and it can be regarded as linear under certain assumptions~\cite{Ganu2013nplug, Baran1989Optimal}. %In this section, we will investigate this relationship on a simplified radial distribution model, and in the next section we will propose a method for extracting this mapping function based on the findings presented here.

\begin{figure}[t]
\centering
\includegraphics[trim=1.1cm 0.6cm 1.1cm 0.6cm, scale=0.22]{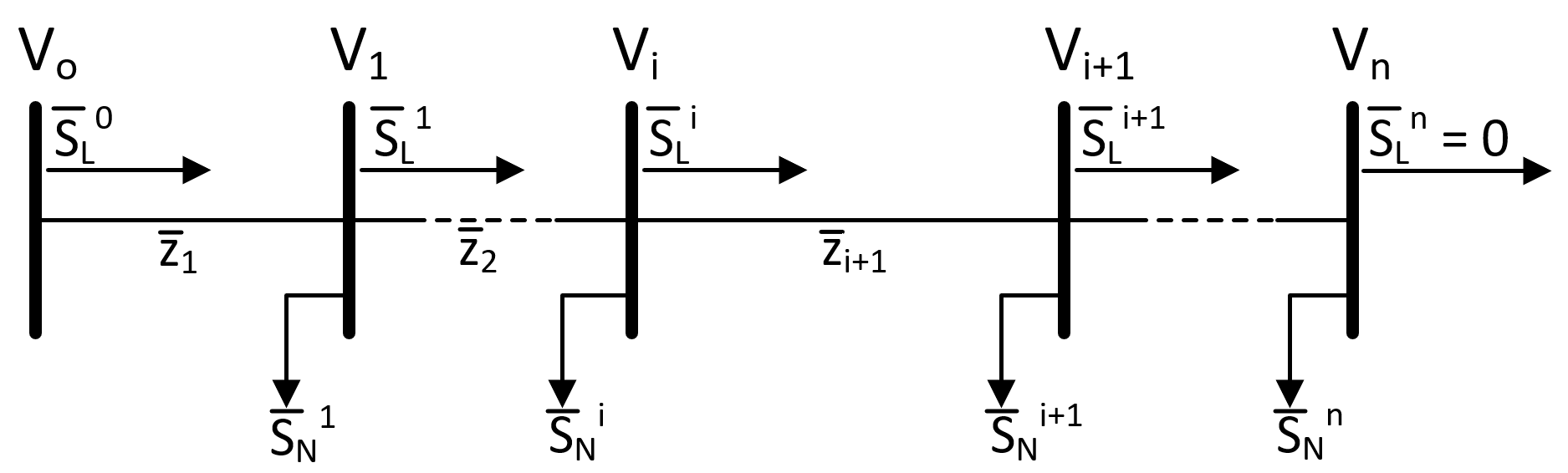}
\caption{Single feeder, radial distribution grid model.}
\label{fig:grid_model}
\end{figure}

We demonstrate this relationship between node voltage and total demand power on a radial distribution grid model shown in Fig.~\ref{fig:grid_model}. In this model, $V_{o}$ represents the substation voltage magnitude, $\{V_{1},...,V_{n}\}$ are magnitudes of node voltages at nodes $1,\ldots, n$, and $\overline{S_L^i} = {P_L^i} + j{Q_L^i}$ (subscript $L$ stands for line) denotes the complex power flowing from node $i$ to node $i+1$ over a line impedance of $\overline{Z_{i}} = r_{i} + jx_{i}$. $\overline{S_N^i} = {P_N^i} + j{Q_N^i}$ (subscript $N$ stands for node) is the complex power drawn from node $i$. This model can be solved for an end-node voltage $V_{i}$ using the distribution grid power flow (\textit{DistFlow} \cite{Baran1989Optimal}) equations (\ref{eq:DistFlow}) shown below:
\begin{equation} \label{eq:DistFlow}
\begin{split}
    {P_L^{i+1}} &{=}{P_L^i}{-} r_{i+1}\frac{{P_L^i}^{2}{+}{Q_L^i}^{2}}{V_{i}^{2}}{-}{P_N^{i+1}}, \\
    {Q_L^{i+1}} &{=}{Q_L^i}{-} x_{i+1}\frac{{P_L^i}^{2}{+}{Q_L^i}^{2}}{V_{i}^{2}}{-}{Q_N^{i+1}}, \\
     V_{i+1}^{2} &{=}V_{i}^{2}{-}2(r_{i+1}{P_L^i}{+} x_{i+1}{Q_L^i}){+}(r_{i+1}^{2}{+} x_{i+1}^{2})\frac{{P_L^i}^{2}{+}{Q_L^i}^{2}}{V_{i}^{2}}.
\end{split}
\end{equation}

By further simplifying the equations using the \textit{LinDistFlow} approximation~\cite{Baran1989Optimal}, we can derive an expression for the voltage of $i^{th}$ node by recursively solving \eqref{eq:DistFlow}
\begin{equation} \label{eq:recursive_voltage}
     V_{i+1}^{2} = V_{i}^{2}{-}2(r_{i+1}{P_L^i}{+} x_{i+1}{Q_L^i})
\end{equation}
where ${P_L^{i}}{=}{P_L^{i-1}}{-}{P_N^{i}}$ and ${Q_L^{i}}{=}{Q_L^{i-1}}{-}{Q_N^{i}}$. Then the final relationship between end-node voltage ($V_{i}$) and total feeder active and reactive powers ($P_{L}^{0}$ and $Q_{L}^{0}$) can be formulated as
\begin{equation} \label{eq:v_vs_power_2}
    V_{i}^{2}{=}V_{0}^{2}{-}{2}P_{L}^{0}\left(R-\mathbb{P}\right)-{2}Q_{L}^{0}\left(X-\mathbb{Q}\right)
\end{equation}
where $R=\sum_{k=1}^{i} r_{k}$, $X=\sum_{k=1}^{i} x_{k}$, $\mathbb{P}=\boldsymbol{\phi^{P}Ur^{T}}$, $\mathbb{Q}=\boldsymbol{\phi^{Q}Ux^{T}}$, $\boldsymbol{\phi^{P}}=[\phi_{1}^{P} \ \phi_{2}^{P} \cdot\cdot\cdot \phi_{i-1}^{P}]$, $\boldsymbol{\phi^{Q}}=[\phi_{1}^{Q} \ \phi_{2}^{Q} \cdot\cdot\cdot \phi_{i-1}^{Q}]$, $\phi_{i}^{P} = \frac{P_{N}^{i}}{P_{L}^{0}}$, $\phi_{i}^{Q} = \frac{Q_{N}^{i}}{Q_{L}^{0}}$, $\boldsymbol{r^T}=[r_{2} \ r_{3} \cdot\cdot\cdot r_{i}]^{T}$, $\boldsymbol{x^T}=[x_{2} \ x_{3} \cdot\cdot\cdot x_{i}]^{T}$, and $\boldsymbol{U}$ is an upper triangular matrix of size $(i-1){\times}(i-1)$ with all the nonzero elements equal to one. $\boldsymbol{\phi^{P}}$ and $\boldsymbol{\phi^{Q}}$ are power consumption ratios that define the percentage of the active and reactive power drawn by each node to the total feeder power ($P_{N}^{i}/P_{L}^{0}$ and $Q_{N}^{i}/Q_{L}^{0}$), and they are stochastic variables by nature. \eqref{eq:v_vs_power_2} reveals that the end-node voltage magnitude square ($V_{i}^2$) is linearly proportional to total feeder power consumption ($P_{L}^{0}$ and $Q_{L}^{0}$). %Since $V_{i}$ is typically allowed to change over a small range (0.9--1~pu), this quadratic equation can even be approximated to a linear one. 

Utilizing the historical local voltage data ($V_{i}$) and total feeder power data ($P_{L}^{0}$ and $Q_{L}^{0}$), the problem of estimating grid loading level turns into a supervised learning problem. This is equivalent to mapping local voltage through a function $\boldsymbol{f}$ to substation powers such that $f(V_{i}) \to P_{L}^{0}, Q_{L}^{0}$. Our goal is to learn this mapping function $f(V_i)$ from historical data. One problem is that $V_{i}$ is coupled with a linear combination of $P_{L}^{0}$ and $Q_{L}^{0}$ through \eqref{eq:v_vs_power_2}, not each of them individually. This requires two mappings from $V_{i}$ to both $P_{L}^{0}$ and $Q_{L}^{0}$. However, in practice, total reactive power consumption $Q_{L}^{0}$ in a distribution power grid is relatively much smaller compared to active power consumption $P_{L}^{0}$, and thus total apparent power $S_{L}^{0}{=}|\overline{S_L^0}|$ can be assumed to be very close to $P_{L}^{0}$. Therefore, we can construct the mapping from $V_{i}$ to $S_{L}^{0}$ such that $f(V_{i}) \to S_{L}^{0}$. This assumption reduces the mapping output to a single variable ($S_{L}^{0}$).% and also takes the reactive consumption into account.

%Therefore, learning voltage-demand lines can be performed when the grid is lightly loaded, and the current deviations are relatively low. %An important conclusion is that nodes can estimate the total demand of the substation by observing the voltage if they can acquire and store their associated voltage-demand lines. We should also note that these lines were obtained at 0\% EV penetration. With higher penetration levels, the distribution grid will be loaded more, and the currents drawn from each node will also have higher deviations, which would adversely affect lines' linearity. To that end, learning voltage-demand lines should better be performed for times during which the grid is lightly loaded, and the current deviations are relatively low.
% \begin{figure}[t]
% \centering
% \includegraphics[trim=0.2cm 0cm 0cm 0cm,clip, scale=0.3]{figures/voltages_and_demand_power.png}
% \caption{Voltage and total demand power data. The blue, yellow and orange waveforms are the voltages of three randomly selected houses whereas the purple waveform represents the total demand power in MVA.}
% \label{fig:voltage_and_demand_power}
% \vspace{-0.2cm}
% \end{figure}

However, there are several factors that disturb this relationship in \eqref{eq:v_vs_power_2} and its parameters. First, $V_{0}$ will have some variations throughout the day %affecting \eqref{eq:v_vs_power_2} 
even though it is usually highly regulated at the feeder level. Second, there are a number of voltage regulation devices such as load tap changers (LTCs), voltage regulators (VRs), and capacitor banks operating at different points in the network. They can be very effective in changing the end-node voltage and thus affect the relationship. Third, distribution grid topology can change due to various reasons including system reconfiguration and expansion. This results in changes in $\boldsymbol{r^{T}}$ and $\boldsymbol{x^{T}}$. Fourth, there are also stochastic variables ($\boldsymbol{\phi^{P}}$ and $\boldsymbol{\phi^{Q}}$ vectors) in \eqref{eq:v_vs_power_2} that change both in space and time.
%, and disturb the relationship. 
Fifth, significant reactive power consumption and generation will also deteriorate the relationship impacting the parameters. All these factors make the problem very challenging and therefore require a rigorous analysis. However, once the modeling and the relationship in \eqref{eq:v_vs_power_2} are resolved, this can truly unlock the ability of end-nodes to probe the grid on their own using local measurements. At this point, high function approximation capacity of ML becomes important. Data is the key to extract these hidden structures within the physical system and their associated time dependencies with the aid of ML.
%\hl{This study does not consider the above factors that can introduce variations and impact the relationship parameters. This will be the subject of a future work where we will use more data that includes more features such as frequency, weather, and power injection or consumption to truly understand the source of variations and train our ML network accordingly.}   

\vspace{-2mm}
\section{AIMD-Based EV Charging Algorithm} \label{sec:proposed_algorithm}
\begin{figure}[t]
\centering
\includegraphics[scale=0.4]{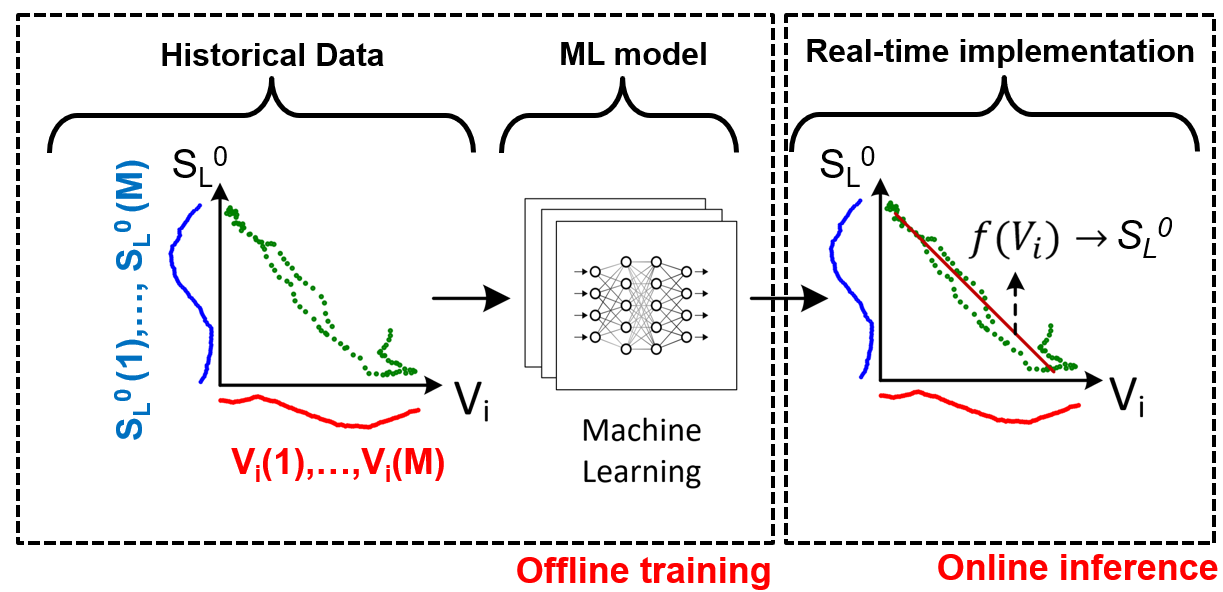}
%\vspace{-2mm}
\caption{Training and implementation phases of the ML network using historical voltage and substation apparent power data and. }
\label{fig:learning_schemetic}\vspace{-0.5cm}
\end{figure}

The MD phase of the AIMD algorithm is triggered by a CE, where the shared capacity reaches its limit. In an ideal implementation of AIMD, every user is notified of CE, and they execute the MD phase of the algorithm reducing the total load simultaneously. Broadcasting CE information to every user requires constant and uninterrupted network connectivity, which increases the communication overhead and makes the system more vulnerable to cyber-attacks. A key goal in this study is to reduce the need for connectivity and operate the system without any information exchange as long as possible. 

Following Section~\ref{section-3}, %we demonstrated that voltage-demand line can be obtained if the substation power profile is known. 
via providing substation loading information to end-users, every user can learn a voltage-substation loading relationship by building an ML regression model locally. {For this reason, the historical substation loading data is made to be accessible to every end-node via a communication network, and end-nodes can download this data only when they need to train their ML model. We note that this is not a real-time feedback signal for our controller. Rather, it is historical power measurements collected at the substation feeder level. The transfer of this data to end-nodes is a unidirectional data flow from the substation to end-nodes, and it is not meant to be real-time. In fact, this information flow can take place as few as once per week or per month if the grid parameter distributions in \mbox{\eqref{eq:v_vs_power_2}} do not shift much. The infrequent and unidirectional information flow of the proposed method reduces the communication dependency of the proposed method making it less vulnerable to cyber-security issues.}
This is illustrated in Fig.~\ref{fig:learning_schemetic}. End-users train their ML models where the inputs to the ML model are time-series local voltage measurements ($\mathbf{V_{i}}$) and the outputs (labels) are time-series total substation feeder apparent power measurements ($\mathbf{S_{L}^{0}}$).

Since \eqref{eq:v_vs_power_2} is a quadratic equation, we can use a second order model for our ML regression network to extract the voltage and total apparent power relationship based on our findings in Section~\ref{section-3}. This model defines a second order polynomial that can be fit to the data by:
\vspace{-1mm}
\begin{equation}\label{eq:V_and_S_vectors}
   {S_{L}^{0}(t)} = \theta_{i,1} + \theta_{i,2}{V_{i}(t)} + \theta_{i,3}{V_{i}^{2}(t)}
    %V_{i} = a_{i} + b_{i}D
    \vspace{-1mm}
\end{equation}
where the subscript $i$ denotes the $i^{th}$ end-node, and $\theta_{i,1}$ $\theta_{i,2}$, and $\theta_{i,3}$ are the polynomial coefficients for the $i^{th}$ node.

% \begin{equation}\label{eq:ML}
%     \mathcolorbox{yellow} {\mathbf{V_{i}}{=}[V_{i}(0)\cdots V_{i}(T)], 
%      ~\text{and}~\mathbf{S_{L}^{0}}{=} [S_{L}^{0}(0)\cdots S_{L}^{0}(T)].}
%     \vspace{-1mm}
% \end{equation}

We will utilize the time-series end-node voltage measurements ($\mathbf{V_{i}}{=}[V_{i}(1)\cdots V_{i}(M)]^{T}$) and time-series substation apparent power measurements ($\mathbf{S_{L}^{0}}{=} [S_{L}^{0}(1)\cdots S_{L}^{0}(M)]^{T}$) to estimate the coefficient vector $\bm{\theta_{i}}=[\theta_{i,1},\theta_{i,2},\theta_{i,3}]^{T}$ for the duration $[1 \ldots M]$. We are investigating the dependencies of two variables (voltage at node $i$ and total apparent power at the substation) and aim to learn a second-order model to build the relationship between the two. In other words, our model has one independent variable, which is the end-node voltage $V_{i}$, and one dependent variable, which is the total substation power $S_{L}^{0}$. Therefore, our model is a bi-variate model even though the actual power grid is a multi-variate system with many dependencies. This can reduce the ability to capture variations in other system parameters, and thus be considered the weakest point of the model. However, our approach assumes that the variations in the other system parameters are insignificant, which is likely valid especially in the short term. We also plan to include more variables (features) in our model to make it more robust to disturbances in a future study.

The three coefficients of the model can be estimated using the least squares regression (LSR). We define the following LSR formulation.
\begin{equation} \label{eq:LSR_formulation}
    \mathbf{S_{L}^{0}} = \mathbf{\Lambda_{i}} \bm{\theta_{i}}
\end{equation}

where $\mathbf{S_{L}^{0}}$ is the substation apparent power measurements, $\bm{\theta_{i}}$ is the coefficient vector, and $\mathbf{\Lambda_{i}}$ is an $M\times3$ matrix that is made of voltage measurements of the $i^{th}$ node and defined as follows
\begin{equation}\label{eq:LSR}
    \mathbf{\Lambda_{i}}=
 \begin{bmatrix} 1 & V_{i}(1) & V_{i}^{2}(1) \\ 1 & V_{i}(2) & V_{i}^{2}(2) \\ \vdots & \vdots & \vdots \\1 & V_{i}(M) & V_{i}^{2}(M)  \end{bmatrix}
\end{equation}
The estimate of parameters ${\bm{\hat{\theta}_{i}}}$ can now be calculated using the closed form solution of the least squares problem as in \eqref{eq:LSR}:
\begin{equation} \label{eq:LSR}
    \bm{\theta_{i}}=(\mathbf{\Lambda_{i}}^{T}\mathbf{\Lambda_{i}})^{-1}\mathbf{\Lambda_{i}}^{T}\mathbf{S_{L}^{0}}
\end{equation}

\begin{algorithm}
\caption{AIMD algorithm for EV charging network}\label{alg1}
\hrule
 \vspace{1mm}
\begin{algorithmic}[1]
 \item[\textbf{Input:}]Substation rated capacity: $S_{L_{(rated)}}^{0}$
 \item[\textbf{Input:}]Voltage meas. : $\mathbf{V_{i}}=[V_{i}(1)\cdots V_{i}(M)]$
 \item[\textbf{Input:}]Substation power meas. : $\mathbf{S_{L}^{0}}=[S_{L}^{0}(1)\cdots S_{L}^{0}(M)]$
 \item[\textbf{Compute:}]  $\bm{\theta_i}$ using \mbox{\eqref{eq:LSR}}
 \item[\textbf{Compute:}]$V_{i,th}$ using \eqref{eq:threshold}
 \item[\textbf{Parameter:}] Additive parameter: $\alpha_{i}=1$
 \item[\textbf{Parameter:}] Multiplicative parameter: $\beta_{i}=0.5$
  
 \item[\textbf{Input:}]Previous charging current: $I_{i}(t)$
 \item[\textbf{Input:}]Node voltage: $V_{i}(t)$
 \item[\textbf{Output:}] New charging current: $I_{i}(t+1)$

\WHILE{SOC $<$ 100\%}
 \IF{$V_{i}(t) > V_{i,th}$ \AND $V_{i}(t) > V_{min}$}
 \STATE $I_{i}(t+1)=I_{i}(t)+\alpha_{i}$
 \ELSE
 \STATE $I_{i}(t+1)=\beta_{i}\times I_{i}(t)$ 
 \ENDIF
 \ENDWHILE
 \vspace{1mm}
\hrule
\end{algorithmic}
\end{algorithm}
 \vspace{-3mm}

As we will demonstrate in the next section, the coefficient of the second order term usually turns out close to zero suggesting that a first order model could have also be used. The reasons a second-order method was preferred in this study are because (1) we tried to build a more generic model that complies with the derivations in Section-III regardless of the nominal voltage range, and (2) there is no additional significant cost of using a second order model in terms of computation. Besides, we also tested a linear regression (LR) model in one of our previous work~\mbox{\cite{Ucer2020Machine}}.

With this ML model, end-users can estimate how much the substation is loaded by only using their local voltage levels. Similarly, users can calculate a particular threshold voltage value ($V_{i,th}$) by solving \eqref{eq:V_and_S_vectors} for voltage given the estimated polynomial coefficients ($\bm{\theta_{i}}$) and substation rated capacity ($S_{L(rated)}^{0}$) such that
\begin{equation} \label{eq:threshold}
    S_{L(rated)}^{0}-\theta_{i,1}-\theta_{i,2}V_{i,th}-\theta_{i,3}V_{i,th}^2 = 0.
\end{equation}
The MD phase of AIMD is triggered if the measured voltage $V_{i}(t)$ drops below $V_{i,th}$. If $V_{i}(t)$ happens to be less than the minimum utilization voltage $V_{min}$, 
then MD should also trigger again. %This prevents voltage quality violation. 

% \begin{figure}[t]
% \centering
% \includegraphics[trim=1.1cm 0.6cm 1.1cm 0.6cm, scale=0.17]{figures/linear_fit_rev2.png}
% \vspace{-3mm}
% \caption{Voltage vs. demand data and the linear line fit with its analytical equation of a randomly selected house.}
% \label{fig:linear_fit}
% \vspace{-0.5cm}
% \end{figure}

The proposed charging algorithm implementing AIMD control using ML-based learned thresholds is presented in Alg.~\ref{alg1}. The algorithm checks for any CE by monitoring the end-node voltage ($V_{i}(t)$) and comparing it with the threshold voltage ($V_{i,th}$) and minimum voltage ($V_{min}$). If the end-node voltage drops below either of these, it triggers the MD phase of AIMD. The period of CE check is defined as the algorithm period $T_{a}$ (not shown in Alg.~\ref{alg1}), and it decides how often the algorithm checks for the triggering conditions. We set it to 10~s considering the power distribution grid and EV charger dynamics. A faster response would cause a more oscillatory behaviour. %The algorithm requires the substation power profile to correlate with local voltage. 
We assume that the voltage is being recorded at high granularity at end-nodes in EV supply equipment (EVSE). The substation power data, on the other hand, does not have to be as granular. %The resolution of the voltage data could be adjusted by down-sampling or time averaging to match that of the power data.%, and having a flexibility over the resolution of the voltage data is only possible if its original sampling rate is high enough (high granularity).
The algorithm needs connectivity only when it fetches the substation data. If we assume that the grid topology or rated capacity does not frequently change, the algorithm can autonomously operate for a long period of duration without exchanging any information. Every node can preserve its voltage-total power characteristics and continue using it until the grid is subject to a significant change/update in terms of supply capacity, loading level, or physical architecture. {Distribution shift is also another problem of data-driven methods, which is discussed in detail in \mbox{\cite{Dobbe2020Learning}}. When there is a distribution shift in the stochastic parameters (e.g., EV charging needs, end-node power profiles, substation voltage, etc.) due to changes in seasons or other important events such as the one we experienced in COVID-19 pandemic, the algorithm will need to understand the changes via more frequently collecting data from the substation and comparing them to the past observations. Each learning phase will update the model parameters to the latest changes. Even though the data distribution is likely to change, the model itself is highly shaped by the topology parameters, which may not change so dramatically in the short term.} The frequency of data exchange greatly depends on these changes and dynamics. End-nodes can decide themselves when they need to re-learn their new voltage-power models by taking longer timescale measurements and making comparisons, which are not studied here.

\vspace{-2mm}
\section{Comparative Performance Evaluation}
%\section{Testing Environment and Performance Metrics} 
\label{section:other_algorithms}

\subsection {Description of the Test System} \label{section:system_description}

\begin{figure}[tb]
  \centering
	\includegraphics[width=86mm]{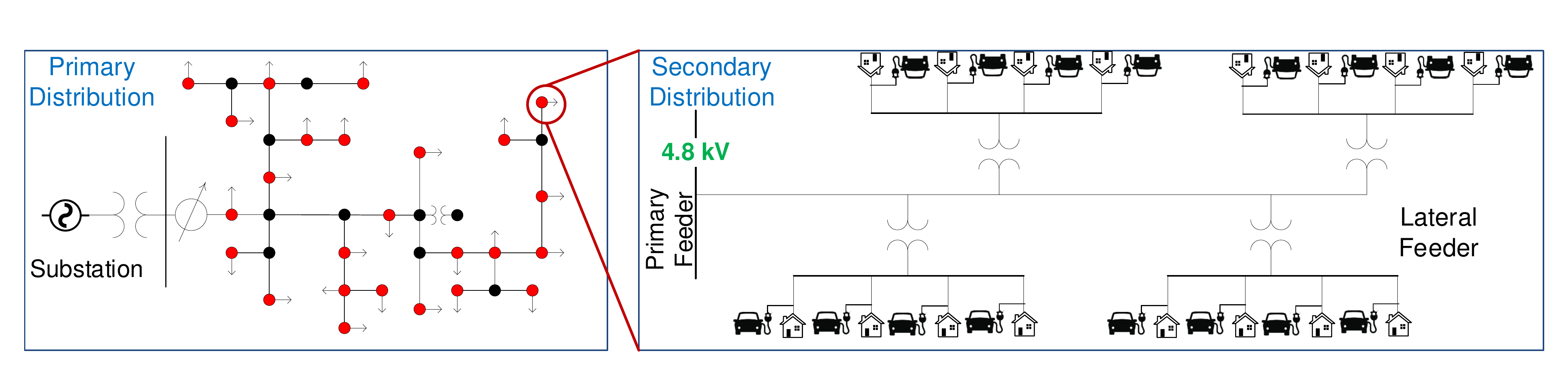}
	\caption{Primary and secondary distribution network implemented in the MATLAB model.}
    \label{fig:grid}
    \vspace{-4mm}
\end{figure}

The simulations are performed in IEEE-37 test distribution grid illustrated in Fig.~\ref{fig:grid}. Each red circle in the figure represents a neighborhood that is connected to the primary network. There are a total of 26 neighborhoods each of which is modeled as a secondary network following a similar procedure described in \cite{malekpour2015radial}.  Each neighborhood contains four 25~kVA transformers ($26\times4=104$ transformers in total) powering four inner nodes. The transformers step down the primary feeder voltage of 4.8~kV to a secondary voltage level of split-phase 120/240~V. Each inner node consists of 4 houses making a total of 16 houses in a neighborhood, and a total of 416 houses in the overall distribution model. Each house is modeled as two separate nodes; one node is dedicated only to EV connection and the other node is reserved for the uncontrollable household load. Hence, there are $416\times2 = 832$ separate nodes in the distribution network. Each end-node represents a residential customer that has a unique household power consumption profile and an EV that is modeled separately. The capacity of the substation transformer is rated at 2.5~MVA, and the grid operates slightly over 1.36~MVA at peak hours without any EV charging event, i.e. the base load.

\begin{figure}[t]
\centering
\includegraphics[trim=1.1cm 0.6cm 1.1cm 0.6cm, scale=0.3]{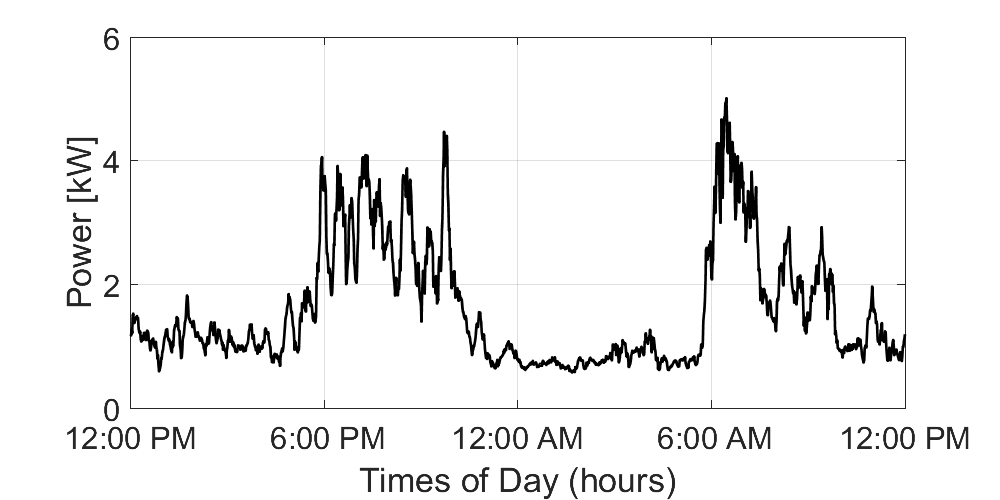}
\caption{A sample household power consumption profile.}
\label{fig:power_profile}
\vspace{-0.6cm}
\end{figure}

For modeling residential loads, we developed a random household load generator. The load profile of each house is generated using 16-day of power consumption data of a real household collected using e-Gauge~\cite{Egauge}. A new power value is generated for each one-min interval from a Gaussian distribution with a mean and standard deviation computed based on 16 data points obtained from the 16-day data. We also assumed a power factor of 0.9 lagging for each house. This assumption is based on our observations on the collected reactive power data \cite{ResidentialDataset2020}. A sample power consumption profile created for one house with this method can be seen in Fig.~\ref{fig:power_profile}. 
The model was developed and simulated in MATLAB Simulink at a time-step of one second, and the simulation takes around 7 mins on a computer with a 3.30 GHz Intel(R) Xeon(R) processor and 16 GB of RAM. The simulations were run from 4:00PM to 11:59PM during when the majority of residential charging takes place. Finally, the same household consumption and EV arrival/departure time data sets were used in each simulation for a fair comparison. Each EV is assumed to have a battery capacity of 72~kWh with an on-board charger of 10~kW. %corresponding to 41~A of AC current at a rated voltage of 240~V. 
We modeled the EVs with the same configuration to prevent possible confusions that could arise when quantifying the fairness performance of the algorithm. %We generate arrival times based on a normal distribution with mean and standard deviation of $\mu=$17h30m and $\sigma=$0h30m. Our model generates initial state of charge (SOC) values for each EV at the time of grid connection based on a Gaussian daily trip distribution with mean and standard deviation of $\mu=$39.5 km and $\sigma=$15.8 km \cite{Erden2018Adaptive}. 

\begin{figure}[b]
\centering
\includegraphics[width=\columnwidth]{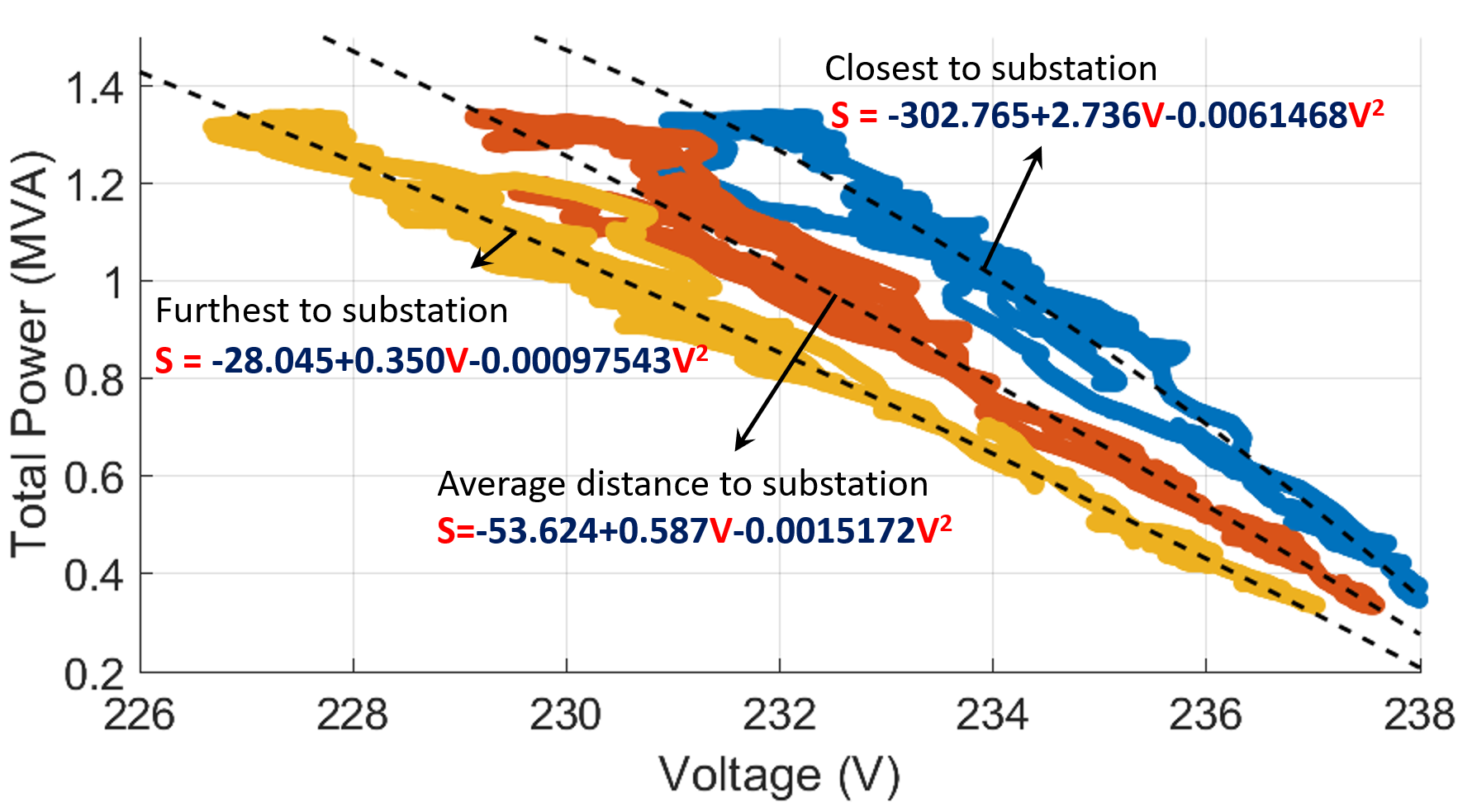}
\vspace{-2mm}
\caption{Voltage vs. total power data for three different houses.}
\label{fig:voltage_vs_demand_power}
\vspace{-0.6cm}
\end{figure}

To verify the relationship between end-node voltage and substation loading, we performed an initial simulation on the developed distribution grid model with no EVs. The voltages of three randomly selected houses as well as the total apparent power consumption of the substation transformer are recorded. Then, the measured voltage samples ($V_{i}$) in one-day are plotted against the corresponding total substation power samples ($S_{L}^{0}$) as shown in Fig.~\ref{fig:voltage_vs_demand_power}. Second-order polynomials fit (dotted lines) to each voltage vs. substation power data as well as their coefficients are also shown. {We can see that the coefficients of $V^{2}$ terms ($\theta_{3}$) are very small compared to other coefficients ($\theta_{1}$, $\theta_{2}$). This suggests that the relationship between voltage and substation power can also be approximated as a linear line. This is mainly because the voltage in a distribution grid is typically allowed to change over a small range (0.9--1~pu).}%This figure clearly shows that a linear relationship can be constructed between voltage and total feeder power especially until the network is loaded less than 0.6 MVA ($\le$25\% of capacity). Beyond this point, the relationship starts deteriorating.

\vspace{-4mm}
\subsection{Performance Score Description} \label{subsec:per_score}

In addition to the well-known metrics of capacity utilization and fairness, we define five more performance scores for fair comparison with the available literature. These scores are $i$) voltage violation, $ii$) global power congestion, $iii$) local power congestion, $iv$) communication overhead, and $v$) average charging power. 

\textbf{Voltage Violation Score (VVS)} is calculated for each algorithm using \eqref{eq:VVS} in V-s as the time integral whose upper bound is the minimum allowed voltage level ($V_{min}=$216~V), and the lower bound is the measured node-voltage $(V_{i}(t))$,
\vspace{-1mm}
\begin{equation} \label{eq:VVS}
VVS =  \frac{1}{N} \sum_{i=1}^{N} \int_{0}^{T}\max\{0, V_{min}-V_{i}(t)\}dt.
\vspace{-1mm}
\end{equation}
where N is the number of nodes, and T is the simulation time horizon, which is 8 hours. If the node voltage never drops below $V_{min}$, then, the score will be zero. 

\textbf{Global Congestion Score (GCS)} is the energy in MVAh consumed beyond the rated substation capacity $S_{L_{(rated)}}^{0}$ for the global congestion. If the power drawn from the substation at time $t$ is denoted by $S_{L}^{0}(t)$, then the GCS can be calculated as in \eqref{eq:GPCS}. 
\vspace{-1mm}
\begin{equation} \label{eq:GPCS}
GCS = \int_{0}^{T}\max\{0, S_{L}^{0}(t)-S_{L_{(rated)}}^{0}\}dt.
\vspace{-1mm}
\end{equation}
If the active power drawn from the substation always stays below the substation's capacity, GCS will be zero.

\textbf{Local Congestion Score (LCS)} is the energy in kVAh consumed over the local distribution transformers' capacity limit. Since there are 104 local transformers in the grid model, the LCS will be presented as an average value for the entire grid and for each 26 neighborhoods. In general, for a distribution system with $K$ distribution transformers with capacity ${S_{{XF_{(rated)}}}^{i}}$ and apparent power {$S_{{XF}}^{i}$} for the $i^{th}$ transformer $(i\in\{1,...,K\})$, LCS can be evaluated as:
\vspace{-1mm}
\begin{equation}\label{eq:LCS}
LCS = \frac{1}{K}\sum_{i=1}^{K}\int_{0}^{T}\max\{0, S_{{XF}}^{i}(t)-{S_{{XF_{(rated)}}}^{i}}\}dt.
\vspace{-1mm}
\end{equation}
where in our case, $K=104$.

\textbf{Capacity Utilization Score (CUS)} is quantified as the maximum percentage (\%) utilization of the capacity as:
\vspace{-1mm}
\begin{equation} \label{eq:CUS}
CUS = \frac{S_{L_{(max)}}^{0}}{S_{L_{(rated)}}^{0}} \times 100\%.  
\vspace{-1mm}
\end{equation}
where ${S_{L_{(max)}}^{0}}$ is the maximum apparent feeder power observed. A higher CUS means a more utilized network system.

\textbf{Average Charging Power Score (ACPS)} is the average of all $N$ charging powers {$(P_{avg}^{i})$} in kW, and calculated as: 
\vspace{-1mm}
\begin{equation} \label{eq:ACPS}
ACPS = \frac{1}{N}\sum_{i=1}^{N}P_{avg}^{i}.
\vspace{-1mm}
\end{equation}

\textbf{Fairness Score (FS)} is calculated based on the \emph{Jain's fairness index} presented in (\ref{eq:fairness_index}), where $\textbf{w}=\{\widetilde{w}^*_{1},\widetilde{w}^*_{2},\cdots,\widetilde{w}^*_{N}\}$ is the vector of average user shares. The same fairness measure is also used in \cite{1989-chiu-analysis}. In the fairest case where everyone has the same exact share, the fairness score becomes 1, whereas in the worst case, the score approaches ${1}/{N}$, where $N$ is the total number of users.
\vspace{-1mm}
\begin{equation} \label{eq:fairness_index}
%\[
FS(\textbf{w}) = FS({\widetilde{w}^*}_{1},{\widetilde{w}^*}_{2},\cdots,{\widetilde{w}^*}_{N}) = \frac{{(\sum {{{\widetilde{w}}^{*}}_{i})}}^{2}}{N\cdot\sum {({{\widetilde{w}}^{*}}_{i})^{2}}}.
%\]
\vspace{-1mm}
\end{equation}
%
%\my{can we normalize Jain's formula by subtracting 1/n from it and then dividing it with (1-1/n)? This way, it will lead to a fairness score between 1 and 0. The formula will look like this: 
% \begin{equation}
% FS(\textbf{w}) = \left( \frac{(\sum w_{i})^2}{\sum w_{i}^{2}} - 1\right) \cdot \frac{1}{n-1}
% \end{equation}

\textbf{Communication Overhead Score (COS)} is based on the number of information exchanges needed for the EVs to implement AIMD. 
\vspace{-3mm}

\subsection{Benchmarks}
%\subsection{Performance Comparison Benchmark Studies}
\label{subsec:other_algorithms}

\begin{figure}[t]
\centering
\includegraphics[trim=1.1cm 0.6cm 1.1cm 0.6cm, scale=0.30]{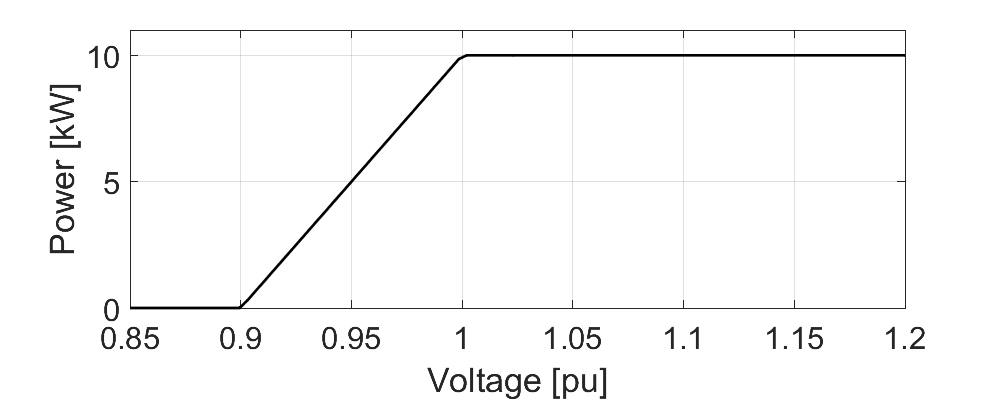}
\caption{{Voltage-charging power droop control function.}}
\label{fig:droop_curve}
\vspace{-0.2cm}
\end{figure}

In order to evaluate the performance of the proposed algorithm, we run the test scenario with two other algorithms presented in \cite{Geth2012Voltage} and \cite{Studli2012AIMD}.%and compare their results with our algorithm using the above selected performance criteria. 
\cite{Studli2012AIMD} proposes an AIMD-based charging control assuming that the capacity CE is notified to each user over a unidirectional communication link in real time. This algorithm is the ideal implementation of the AIMD algorithm and therefore will serve as a benchmark. % for the proposed solution. 
$\alpha$ and $\beta$ parameters presented in Alg.~\ref{alg1} are set to 1 and 0.5, respectively. These values are chosen specifically based on their domain set ($\alpha{>}0$ and $0{<}\beta{<}1$). An increment current of 1~A RMS in each AI phase is assumed.
{In \mbox{\cite{Geth2012Voltage}}, the authors propose to use droop control where the power-voltage droop functions are pre-defined. In this study, we adapt the LM2 type droop function presented in \mbox{\cite{Geth2012Voltage}} for the nominal charging power of 10~kW. The droop function of this controller is illustrated in Fig.~\mbox{\ref{fig:droop_curve}}. 
The algorithm cuts off the charging power when the voltage goes below 0.9~pu (216~V). Charging power is linearly changed with voltage when the voltage is between 0.9-1.0~pu. When the voltage is above 1.0~pu, the EVs will be charged at the maximum rated power of 10~kW.}

\vspace{-3mm}
\section{Simulation Test Results}\label{section:results}
%\mck{In all of the simulation results section presented below, I suggest that we put D-AIMD at the end not to the second place in the order. Otherwise, it is hard to follow what our contribution is.}

We refer the algorithms as follows: \emph{D-AIMD} for our proposed distributed AIMD algorithm, \emph{C-AIMD} for the ideal centralized AIMD algorithm~\cite{Studli2012AIMD}, \emph{Droop} for the droop control algorithm~\cite{Geth2012Voltage}, and \emph{No-Control} for the rated charging power for all EVs. To test the proposed D-AIMD, we first run the simulation with 0\% EV penetration and obtain the substation power data. Then, this data is sent to the end-nodes, and every node computes three coefficients for their voltage-demand polynomial (Fig.~\ref{fig:voltage_vs_demand_power}). Nodes find a threshold voltage using \eqref{eq:threshold} and implement Alg.~\ref{alg1}. We then repeat the simulation at 100\% EV penetration and with all EVs implementing the algorithm. Finally, we compare the performance score of each algorithm. The results are listed in Table~\ref{tab:score_table}.

\begin{table}[tb] 
\caption{Performance comparison with different algorithms.}
\vspace{-2mm}
  \centering
\resizebox{\columnwidth}{!}{  
\begin{tabular}{l|lllllll}
\hline
Algorithm&VVS &GCS &LCS &CUS &ACPS &FS&COS \\
&(V-s)&({MVAh)} &({kVAh}) &(\%) &(kW)& &\\
 \hline
 \hline
No-Control& 21,938 & 1.82 & 64.80 & 170.8 & 8.91 & 0.999 & 0 \\
D-AIMD & 0 & 0 & 4.34 & 99.96 & 4.78 & 0.944 & 1 \\
C-AIMD~& 0 & 9.34$\times10^{-4}$ & 15.87 & 100.47 & 4.95 & 0.958 & 2880 \\
Droop& 0 & 0 & 0.183 & 93.16 & 3.25 & 0.980 & 0 \\
 \hline
\end{tabular} 
}
\label{tab:score_table}
\vspace{-4   mm}
\end{table}

\vspace{-3mm}
\subsection{Voltage Stability vs. Capacity Utilization}

\begin{figure}[b]
\vspace{-3mm}
\centering
\includegraphics[scale=0.18]{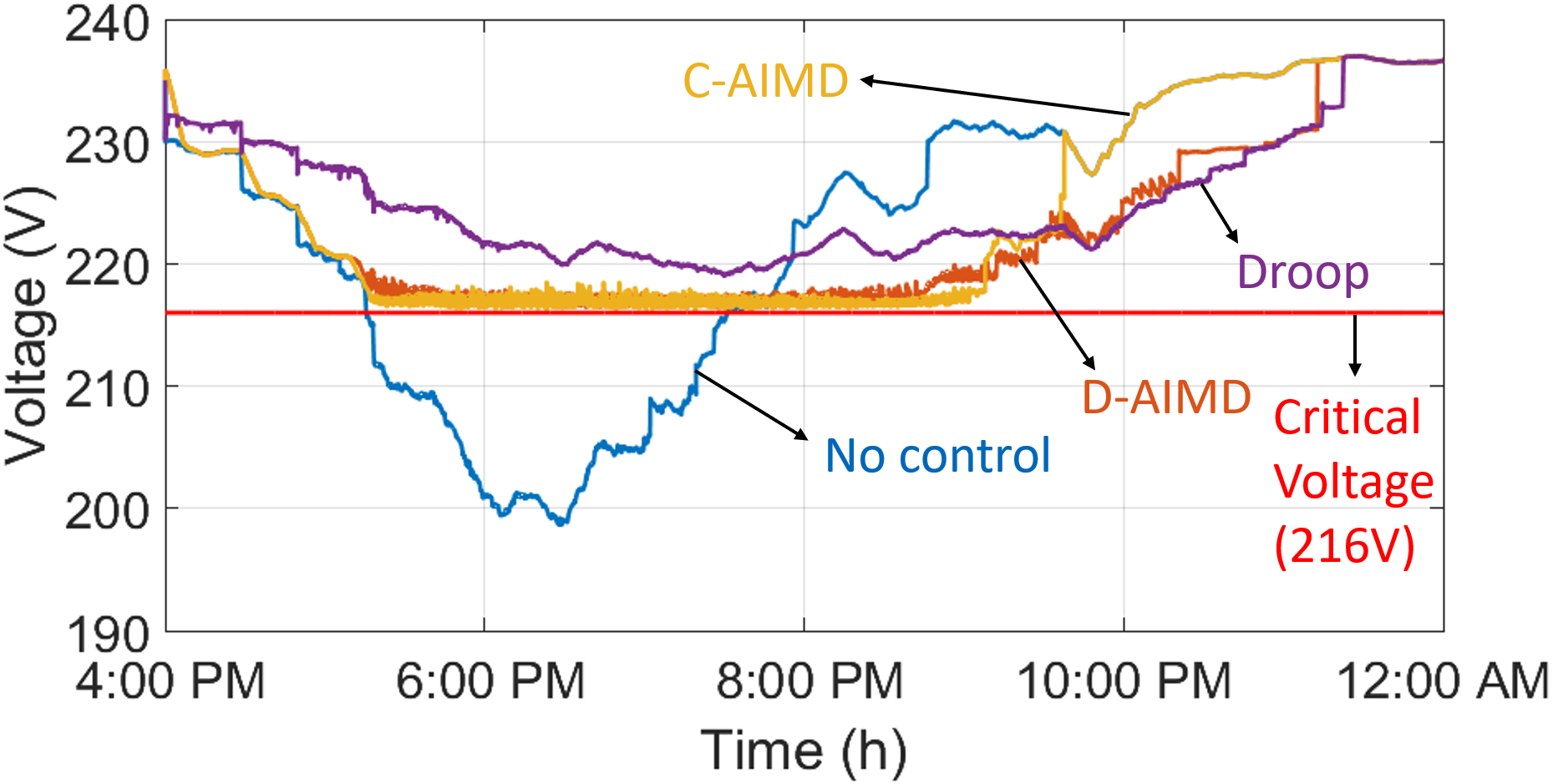}
\vspace{-7mm}
\caption{Minimum voltage waveforms for all three algorithms.}
\label{fig:minimum_voltage}
\vspace{-3mm}
\end{figure}

AIMD-based control successfully avoids voltage violations across the grid since its triggering condition also includes minimum voltage check. Fig.~\ref{fig:minimum_voltage} shows the minimum of voltage values measured across all 416 end-nodes at every second for each algorithm. Without any proper control (i.e., No-control case), the minimum voltage drops way below the critical level of 216~V resulting in an average VVS of 21,938 V-s as shown in Table~\ref{tab:score_table}. C-AIMD and D-AIMD algorithms operate close to the critical voltage level since they have a condition that triggers MD phase when voltage limit is about to be violated. Droop algorithm results in a much higher minimum voltage. Consequently, none of the three algorithms violated the voltage limits of the grid, and therefore, their VVS are all zero.

\begin{figure}[t]
\centering
\includegraphics[scale=0.18]{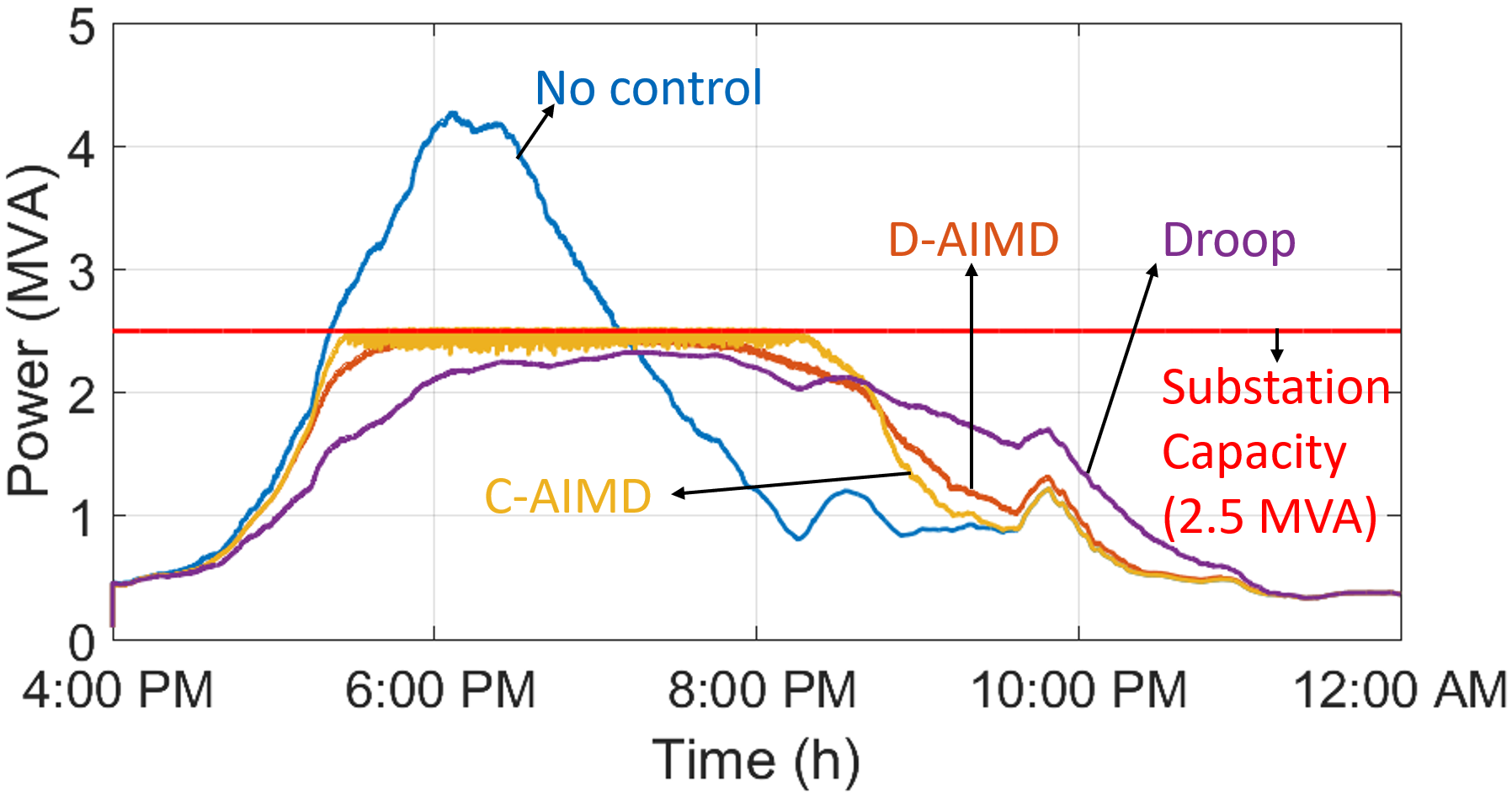}

\caption{Substation transformer apparent powers for all three algorithms.}
\label{fig:substation_power}
\vspace{-3mm}
\end{figure}

Ideally, AIMD ensures the full utilization of system capacity since it is using the capacity congestion as its triggering mechanism. Fig.~\ref{fig:substation_power} presents the resulting substation apparent power for each algorithm. The red horizontal line indicates the substation capacity (2.5~MVA). In No-Control, the capacity is exceeded by more than 70\% with a CUS of 170.8\% and GCS of 1.82~MVAh (Table~\ref{tab:score_table}). C-AIMD triggers when the capacity is reached, and this information is sent to the end-nodes. Therefore, it operates at full capacity (100.47\% in Table~\ref{tab:score_table}). D-AIMD, however, receives no information regarding the capacity, and makes its own estimation based on the derived voltage-power demand polynomial. It still manages to operate near the full capacity (99.96\%), though slightly less than the ideal case (C-AIMD). Droop is utilizing less of the total capacity (93.16\%). %compared to D-AIMD and C-AIMD algorithms. 
The reason that Droop is the least efficient in terms of utilization is that it uses the same droop function for all nodes. This causes the nodes experiencing low voltage to have also low charging power due to the linear droop function. The AIMD-based algorithms, however, keep increasing their charging power until the capacity limit is reached, which allows them to attain a higher utilization. On the positive side for Droop, it stays farther away from the critical voltage. Droop, essentially, provides a coarser trade-off between voltage stability and capacity utilization. Note that it is possible to tune the AIMD-based algorithms to a $V_{min}$ so that the distance to the critical voltage stays high in return of a drop in the capacity utilization. This can be the case when the substation peak loading is to be shaved below its rated capacity.

%And we observe that the substation congestion (global congestion) can be prevented with any of the presented algorithms in action, and therefore their substation congestion scores are zero (Table~\ref{tab:score_table}). However, there is still a significant difference among the results in terms of the capacity utilization, pole transformer loading (local congestion) and communication overhead, which will be discussed next. 

\begin{figure}[tb]
\centering
\includegraphics[scale = 0.33]{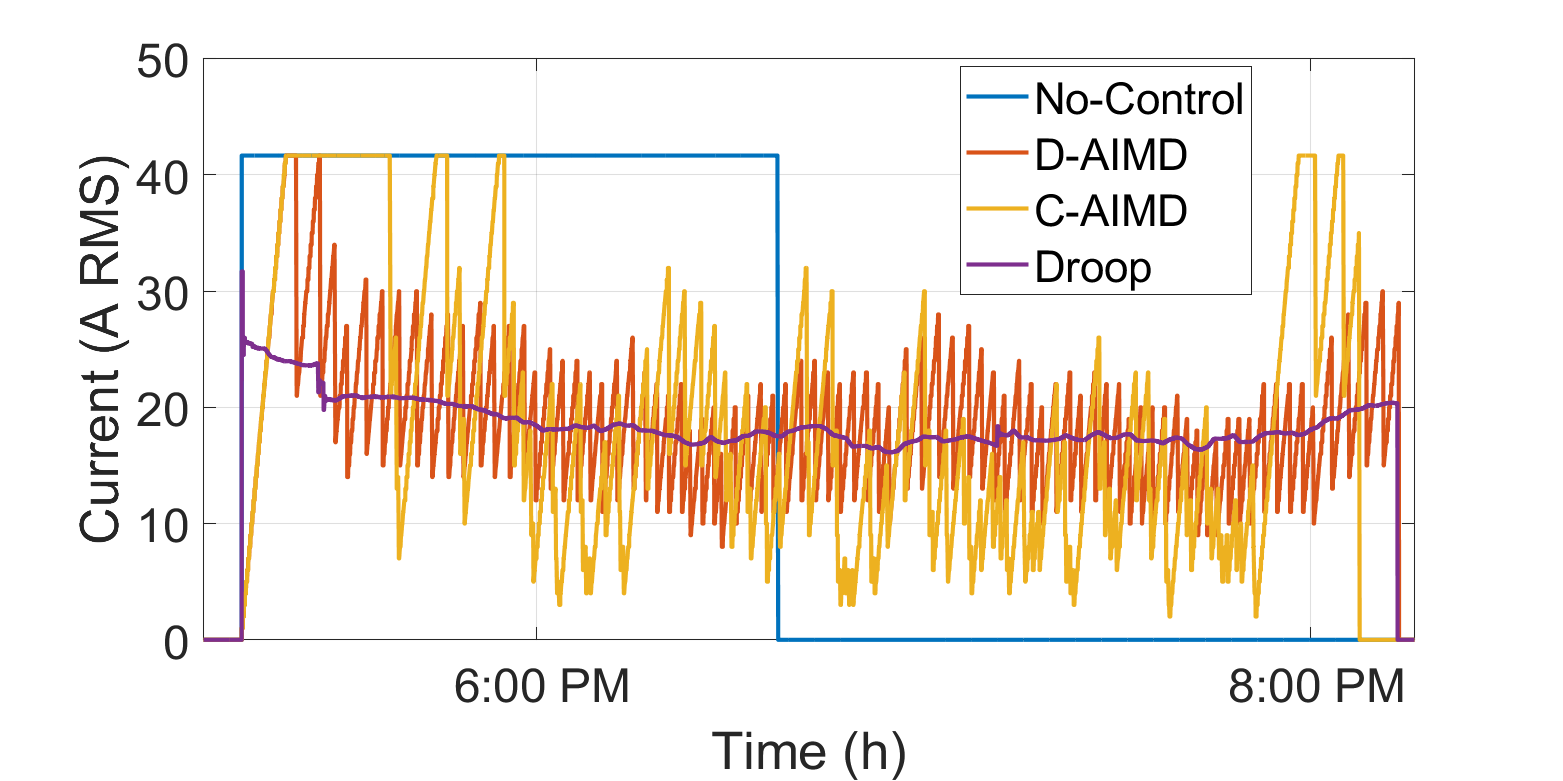}

\caption{Time-domain AC RMS current waveform for all charging algorithms.} 
\label{fig:currents}
\vspace{-0.6cm}
\end{figure}

We present the current waveforms of a randomly chosen EV under four different algorithms in Fig.~\ref{fig:currents}. Since the data-set used in the simulations are the same, the EV's arrival time does not change per algorithm. In No-control, the EV is charged at the fixed current of 41~A until fully charged. D-AIMD and C-AIMD are dynamic in nature (due to AIMD) and constantly increase or decrease the charging current based on the network congestion. The current is more steady in Droop due to its pre-defined linear control droop function.

\vspace{-3mm}
\subsection{Fairness}
\vspace{-1mm}

\begin{figure}[b]
\vspace{-3mm}
\centering
\includegraphics[scale=0.3]{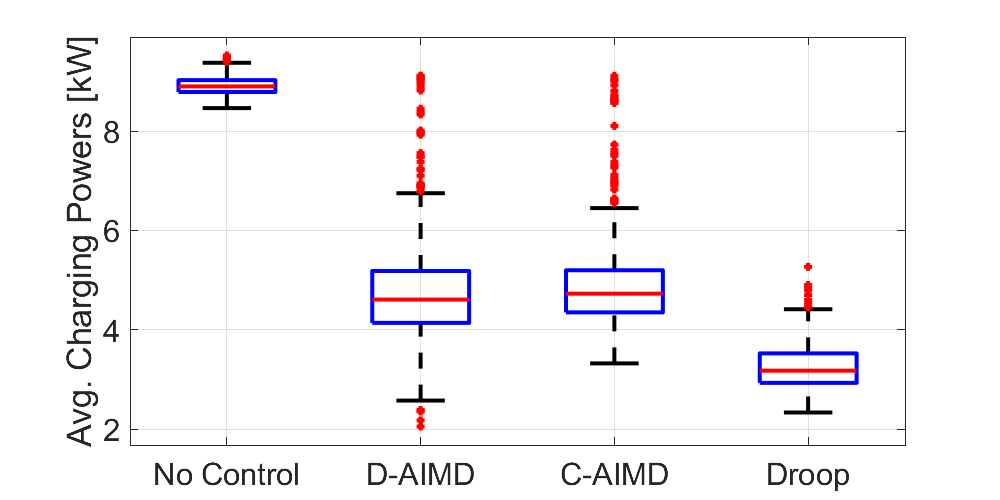}
\vspace{-3mm}
\caption{Distribution of average charging power of EVs for No-Control, D-AIMD, C-AIMD, and Droop algorithms.}
\label{fig:ave_chr_pwrs}
\vspace{-3mm}
\end{figure}

AIMD allocates the available capacity among users in the fairest way if every end-user is ideally notified of the congestion at the same time. Fig.~\ref{fig:ave_chr_pwrs} demonstrates the distribution of average charging powers of 416 EVs for each algorithm in boxplots. No-control naturally results in the highest ACPS of 8.91~kW on average and highest FS of 0.999 since it charges at the maximum charging power for all EVs. %However, there are still slight variations due to the voltages being different all across the grid. 
The charging power distributions of D-AIMD and C-AIMD show that the resulting ACPSs are reduced below 5~kW (4.78 kW for D-AIMD and 4.95 kW for C-AIMD) to alleviate the grid stress. Although both algorithms generated nearly similar distributions, we can infer from the mean and the spread of the distributions that C-AIMD has a slightly higher mean value and less variation, and thus closer charging powers (i.e. fairer charging) due to its centralized mechanism. This leads to the FS of 0.944 for D-AIMD and 0.958 for C-AIMD as listed in Table~\ref{tab:score_table}. Therefore, the proposed D-AIMD has almost the same performance as the ideal (C-AIMD) algorithm without needing any central information, and it starts operating only with local measurements after receiving the power demand information and extracting the voltage vs. power polynomial. Droop algorithm, on the other hand, resulted in the lowest average charging power (3.25 kW) but a narrower distribution spread indicating that it is fairer (0.980) than D-AIMD and C-AIMD. %The variations in charging powers in each algorithm do not occur only because of the algorithms but also due to different arrival times of EVs within each algorithm (i.e. different voltage levels at the time of plug-in). Note that, all four algorithms use the same mobility statistics.% as was explained in Section~\ref{section:system_description}. 

\vspace{-3mm}
\subsection{Congestion}
\vspace{-1mm}

Table~\ref{tab:score_table} shows that 
%the other parts of the grid (neighborhoods) also experience local congestion if No-Control is implemented. 
No-control results in an LCS of 64.80~kVAh. By applying either of the three control algorithms, the substation congestion can be entirely avoided ($GCS\approx0$~MVAh). However, despite its significant reduction, the local congestion is still present. We see that the lowest LCS of 0.183~kVAh occurs with Droop, since its capacity utilization is not as much as the other two. Further, D-AIMD causes less local congestion than C-AIMD (i.e., an LCS of 4.34~kVAh vs. 15.87~kVAh in Table~\ref{tab:score_table}). This is a direct result of higher capacity utilization of C-AIMD. Both AIMD-based algorithms do not directly take local congestion into account, and they operate based on global congestion information. C-AIMD can be greedier and causes more local congestion since it receives the global congestion information directly from the substation whereas D-AIMD acts upon an estimation of congestion at the end-node.

\vspace{-3mm}
\subsection{Communication Overhead}
\vspace{-1mm}

A key parameter to evaluate the algorithms is their need for central information. We defined COS to be the number of information exchanges. No-control and Droop charging require no additional information since they are fully decentralized, and therefore their COSs are both zero in Table~\ref{tab:score_table}. D-AIMD receives substation power demand data only one time before charging and calculates a threshold based on it, which results in a COS of only 1 in Table~\ref{tab:score_table}. C-AIMD, however, needs a central notification of the capacity event to operate. This means that the end-nodes must request this information at every algorithm period ($T_{a}$), which is 10~s. This case study simulates only 8 hours between 16:00-24:00, and the information regarding the presence of the congestion is broadcast at 10~s intervals. Therefore COS for C-AIMD appears to be the highest (2,880) in Table~\ref{tab:score_table}. So, using D-AIMD has significantly reduced the communication overhead and made the control operation almost fully decentralized.  

{We should note that using an external data (e.g. substation loading) certainly increases the amount of information and communication needs compared to fully-autonomous (standalone) solutions. However, our approach targets to minimize these dependencies by only requiring the substation loading, which technically carries no information regarding the grid topology and its parameters, and reducing the frequency of information exchange. This advantage highlights practical advantage of using D-AIMD with more secure and resilient operation against cyber attacks.}

\vspace{-3mm}

\section{Conclusions} \label{section:conclusions}
\vspace{-1mm}
%In this study, we categorize the algorithms used for EV charging based on their operation structure (centralized/decentralized) and need for global/local information. 
In this paper, we introduced the AIMD algorithm that is adapted from the Internet and proposed an ML-assisted, distributed AIMD controller for EV charging. We exploited the relationship between voltage and power demand in a low-voltage distribution grid and presented a mathematical model of this relationship. Based on this relationship, we then proposed an ML-based method to estimate the grid capacity using only local voltage measurements. 

We tested the proposed approach along with two other algorithms proposed in the literature on a test distribution grid. %Then, we did a rigorous comparison including the No-Control case based on certain performance scores. 
We concluded that all algorithms manage to avoid global power congestion and voltage violation. The performance of the proposed algorithm is very close to the ideal AIMD in terms of capacity utilization, average charging power, and fairness. The proposed D-AIMD method outperforms the droop control in capacity utilization by 6.8\%, and in average charging power by 47\% (+1.53 kW). Therefore, D-AIMD uses the available capacity more efficiently and performs charging faster compared to Droop control. As a result, it causes more average local congestion (+4.157 kVAh) and slightly less fair charging (-3.6\%) than Droop control. The proposed control significantly reduces the communication overhead need of the ideal C-AIMD  (from a COS of 2880 to 1) and performs almost the same in terms of fairness and global congestion. In addition, the proposed algorithm outperforms the centralized one with respect to local congestion (from LCS of 15.87 kVAh to 4.34 kVAh).

{We should note that this study does not consider the external factors that may impact the voltage and substation loading relationship. We can only provide some formal guarantees to the system operator that the system operate as expected after truly covering all these possible factors and worst case scenarios. This will be the subject of a future work where we will use more data that includes additional features such as frequency, weather, and end-node power injection or consumption to truly understand the source of variations and train our ML network accordingly. We will also investigate whether it is possible and to what extend to provide these formal guarantees.}

\vspace{-2.5mm}

\bibliographystyle{IEEEtran}
\bibliography{./bibtex/bib/EV_charging}


\ifCLASSOPTIONcaptionsoff
  \newpage
\fi

\vspace{-12mm}

\begin{IEEEbiography}[{\includegraphics[width=1in,height=1.25in,clip,keepaspectratio]{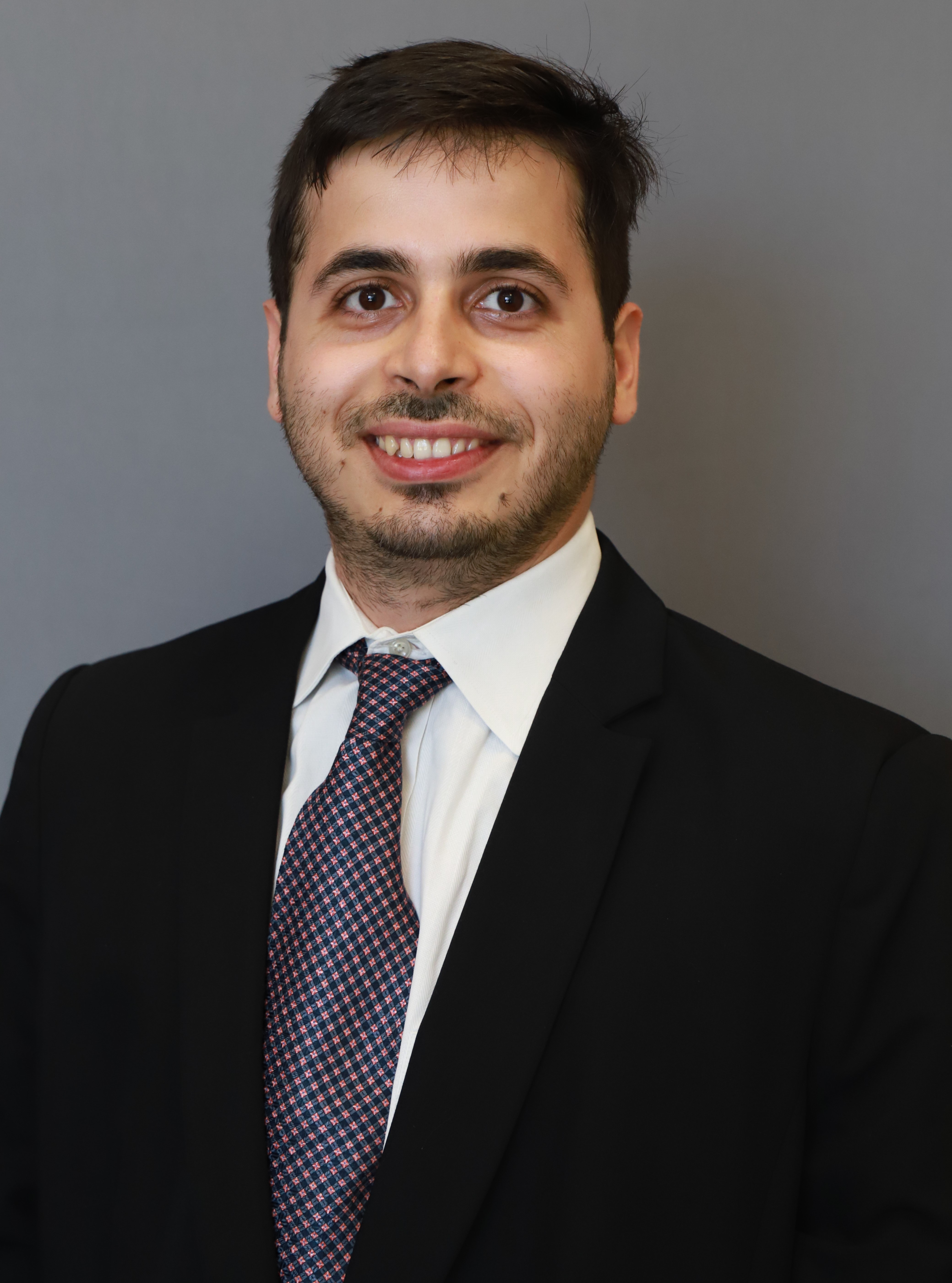}}]{Emin Ucer}(S’18) received the B.S. degree in Electrical and Electronics Engineering from Hacettepe University, Ankara, Turkey, in 2015. He worked as a Research Engineer in TUBITAK between 2015-2016. He’s been working as a Ph.D. student at The University of Alabama since 2017. His research interests are power electronics, electric vehicles (EVs), EV-grid integration and control. He is also interested in machine learning based techniques and their applications.
\end{IEEEbiography}
\vspace*{-3.0\baselineskip}

\begin{IEEEbiography}[{\includegraphics[width=1in,height=1.25in,clip,keepaspectratio]{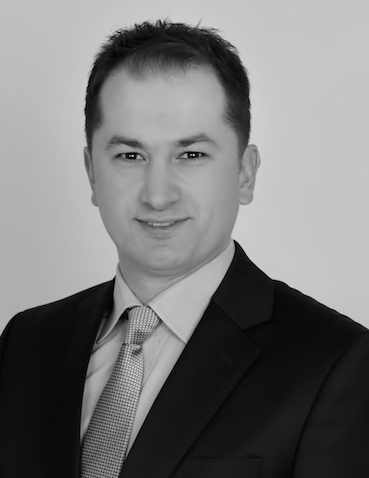}}]{Mithat Kisacikoglu}
(S’04–M’14-SM'21) received the Ph.D. degree from the University of Tennessee in 2013 in electrical engineering. He worked at National Renewable Energy Laboratory, Golden, CO as a research engineer between 2015-2016. He is currently an Assistant Professor in the Electrical and Computer Engineering at The University of Alabama, Tuscaloosa. His research interests include  power electronics converters and EV-grid integration.

Dr. Kisacikoglu is the co-recipient of IEEE PES General Meeting prize paper award in 2019. He is an Associate Editor of The IEEE TRANSACTIONS ON INDUSTRY APPLICATIONS and IEEE TRANSACTIONS ON VEHICULAR TECHNOLOGY. 
\end{IEEEbiography}
\vspace*{-3.0\baselineskip}

\begin{IEEEbiography}[{\includegraphics[width=1in,height=1.25in,clip,keepaspectratio]{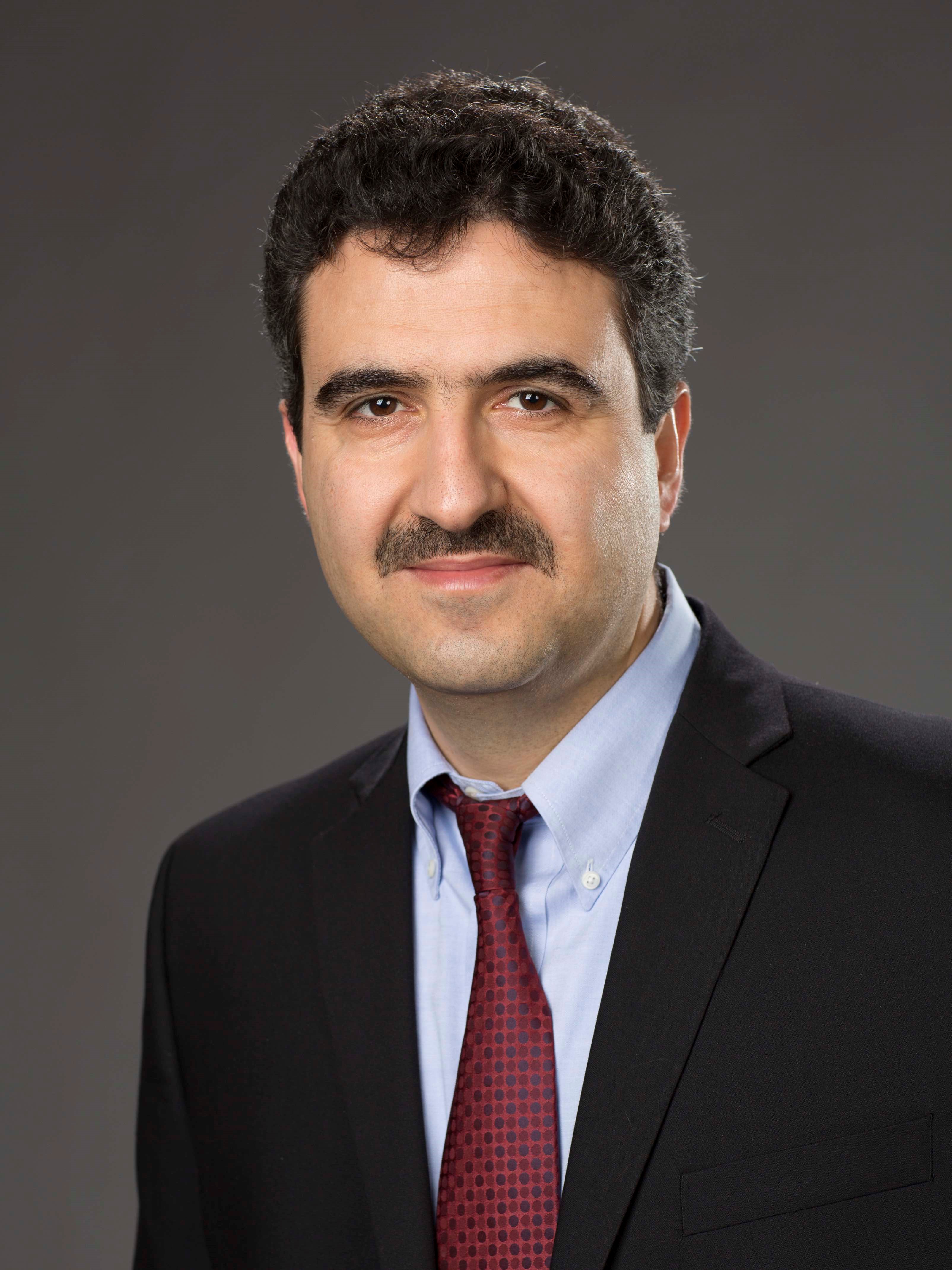}}]{Murat Yuksel}(SM'11) received his Ph.D. degree in computer science from Rensselaer Polytechnic Institute in 2002. He is currently a Professor at the ECE Department of the University of Central Florida (UCF), Orlando, FL. Prior to UCF, he was with the CSE Department of the University of Nevada – Reno as a faculty member until 2016. %He was with the ECSE Department of Rensselaer Polytechnic Institute (RPI), Troy, NY as a Postdoctoral Associate and a member of Adjunct Faculty until 2006. 
His research interests are in the area of networked, wireless, and computer systems. %with a recent focus on big-data networking, UAV networks, optical wireless, public safety communications, device-to-device protocols, economics of cyber-security and cyber-sharing, routing economics, network management, and network architectures. 

He has been on the editorial board of Computer Networks, and published more than 150 papers at peer-reviewed journals and conferences and is a co-recipient of the IEEE LANMAN 2008 Best Paper Award. He is a senior member of IEEE, senior and life member of ACM, and was a member of Sigma Xi and ASEE.
\end{IEEEbiography}
\vspace*{-3.0\baselineskip}

\begin{IEEEbiography}[{\includegraphics[width=1in,height=1.25in,clip,keepaspectratio]{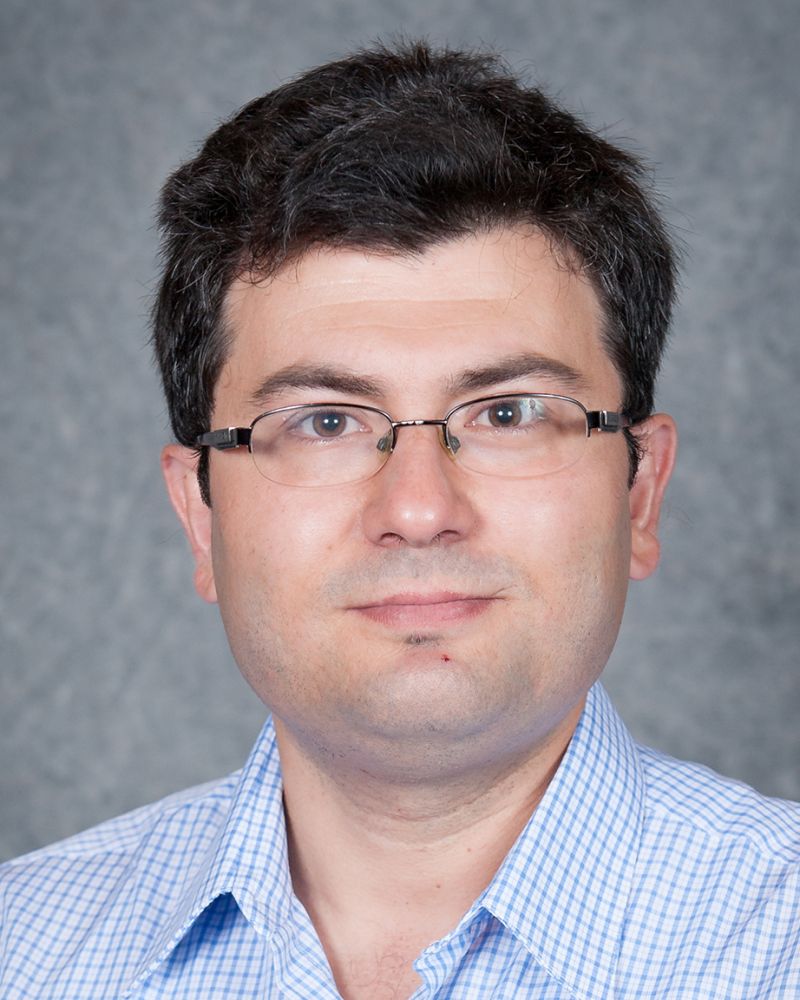}}]{Ali Cafer Gurbuz}(M’08–SM'18)  received the Ph.D. degree from Georgia Institute of Technology in 2008 in Electrical and Computer Engineering. He is currently an Assistant Professor in Electrical and Computer Engineering at Mississippi State University (MSU). %From 2003 to 2009, he researched compressive sensing based computational imaging problems at Georgia Tech.
Prior to MSU, he held faculty positions at TOBB University and The University of Alabama. 
His research areas are machine learning, sparse signal representations, compressive sensing theory and applications, radar and sensor array signal processing.  

He is the recipient of The Best Paper Award for Signal Processing Journal in 2013, the Turkish Academy of Sciences Best Young Scholar Award in Electrical Engineering in 2014 and NSF CAREER award in 2021. He has served as an associate editor for several journals such as Digital Signal Processing, EURASIP Journal on Advances in Signal Processing and Physical Communications. 
\end{IEEEbiography}

\end{document}